\documentclass[preprint,showpacs,preprintnumbers,amsmath,amssymb]{revtex4}
\usepackage{amsmath,amsthm,amssymb}
\usepackage{graphicx}

\begin{document}
\title{Effective Polar Potential in the Central Force Schr\"{o}dinger Equation}
\author{M.S.Shikakhwa\footnote{On sabbatical leave from the Department of Physics, University of Jordan}}
\email{mohammad@metu.edu.tr}
\author{M.Mustafa}
\email{mustafa@purdue.edu}
\address{Physics Program, Middle East Technical University North Cyprus Campus\\
Guzelyurt, TRNC\\
via Mersin 10, Turkey}%

\begin{abstract}
The angular part of the Schr\"{o}dinger equation for a central
potential is brought to the one -dimensional "Schr\"{o}dinger
form" where one has a kinetic energy plus potential energy
terms. The resulting polar potential is seen to be a family of
potentials characterized by the square of the magnetic quantum
number $m$. It is demonstrated that this potential can be viewed
as a confining potential that attempts to confine the particle to
the $xy$-plane, with a strength that increases with increasing
$m$. Linking the solutions of the equation to the conventional
solutions of the angular equation, i.e. the associated Legendre
functions , we show that the variation in the spatial distribution
of the latter for different values of the orbital angular quantum
number $l$ can be viewed as being a result  of "squeezing"  with
different strengths  by the introduced "polar potential".
\end{abstract}
\maketitle
\section{Introduction}
The Schr\"{o}dinger equation for a central potential is a "classic
" example in undergraduate and first year graduate quantum
mechanics courses (see \cite{griffithsQM}, for example). It is a
good exercise in the separation of variables; a good example of
the conservation of angular momentum, and a starting point for
treating the Hydrogen atom problem. An important concept that is
introduced along with this problem is that of the "effective
radial potential", where the central potential that appears in the
radial equation - after separation of variables-acquires and
additional term that depends on the total angular momentum quantum
number $l$ (see below). One then has a one-dimensional
Schr\"{o}dinger equation with an effective potential
\begin{equation}\label{effective}
V_{eff}(r)=V(r) + \frac{\hbar^2}{2m}
\frac{l(l+1)}{r^2}
\end{equation}
The behavior of the radial wave function is interpreted in view of
this effective potential. In particular, the new repulsive part,
is interpreted as a centripetal barrier that attempts to throw the
particle away from the origin, thus determining its shape for
different values of $l$. The present paper analyzes the
 angular equation along similar lines (for the polar variable $\theta$) and
demonstrate that, viewed as a one-dimensional Schr\"{o}dinger
equation, we can identify an effective polar potential that shapes
the angular part of the solution. The aim is to give the students
of quantum mechanics a "feeling" of how the spherical harmonics,
that constitute the angular part of the wave function for any
radial potential, assume the distribution they have about the
$z$-axis. We will identify a polar potential - more precisely, a
family of polar potentials that depend on the magnetic quantum
number $m$ - that squeezes the particle towards the $xy$-plane
with a strength that increases with $m$. The physical picture
introduced, and the intuition employed, we believe, will also
provide the students with a simple prototype of the intuitive
thinking that is employed by practicing physicists in
understanding more complex real-world systems.
\section{The Solutions of the  Schr\"{o}dinger Equation for a Central Potential}
The Schr\"{o}dinger equation
$$
 -\frac{\hbar^2}{2 m} \nabla^2 \psi + V \psi = E \psi
$$
separates in  spherical coordinates for  a central potential by writing $\psi =
R(r) Y(\theta,\phi) $, and introducing the separation constant  $l(l+1)$ . The
resulting radial and angular equations are, respectively:

\begin{equation}\label{radialeq1}
\frac{1}{R} \frac{d}{dr} \left(r^2 \frac{dR}{dr}\right) - \frac{2mr^2}{\hbar^2}
\left[V(r) - E\right] = l(l+1)
\end{equation}

\begin{equation}\label{angulareq1}
\frac{1}{\sin \theta} \frac{\partial}{\partial \theta} \left(\sin \theta
\frac{\partial Y}{\partial \theta} \right) + \frac{1}{\sin^2 \theta}
\frac{\partial^2 Y}{\partial \phi^2} = - l(l+1) Y
\end{equation}

The radial equation is brought to the Schr\"{o}dinger form
 by substituting:

$$u(r) = rR(r)$$

so equation \eqref{radialeq1} becomes:

\begin{equation}\label{radialeq2}
-\frac{\hbar^2}{2m} \frac{d^2 u}{dr^2} + \left\{V(r) + \frac{\hbar^2}{2m}
\frac{l(l+1)}{r^2}\right\} u = E u
\end{equation}

with the  term in the brackets being the effective potential we
mentioned above, equation \eqref{effective}.

Now if we look at equation \eqref{angulareq1}, the so called angular equation,
we can do one more separation of variables by setting
$Y(\theta,\phi)=\Theta(\theta)\Phi(\phi)$, then we will have the azimuthal angle
equation,
\begin{equation}\label{azimuthal}
 \frac{d^2\Phi}{d\phi^2}+m^2\Phi=0
\end{equation}

and the polar equation,

\begin{equation}\label{thetaeq1}
 \frac{1}{\sin \theta} \frac{d}{d \theta} \left(\sin \theta \frac{d\Theta}{d
\theta} \right) - \frac{m^2}{\sin^2 \theta} \Theta = - l(l+1) \Theta
\end{equation}

the solutions of \eqref{azimuthal} are
\begin{equation}
\Phi(\phi)= e^{im\phi}
\end{equation}

and we require the solutions $\Phi(\phi)$ to be single valued, i.e
$$\Phi(\phi+2\pi)=\Phi(\phi)$$
which dictates that $m=0,\pm1,\pm2,...$. It is worth to recall
that the r.h.s of \eqref{angulareq1} is the total angular momentum
hermitian operator $L^2$ represented in spherical coordinates. In
general $L^2$ admits integer and half integer solutions for the
eigenvalue $l$, which is the result of the angular momentum
commutation relations. The restriction of $l$ to integers is a consequence of restricting $m$ to integers.

Let's go back to equation \eqref{thetaeq1}, and write down its
famous solutions which are the associated Legendre functions
\cite{griffithsQM,boas}($N_l^m$ is a normalization constant):

\begin{equation}
 \Theta(\theta)=N_l^m P^m_l(\cos\theta)
\end{equation}

where $P^m_l$ are  given by

\begin{equation}
P^m_l(x) = (1-x^2)^{\frac{|m|}{2}} \left(\frac{d}{dx}\right)^{|m|} P_l(x)
\end{equation}

where $P_l(x)$ are the the Legendre polynomials which are given
by the Rodrigues formula \cite{griffithsQM,boas}:

\begin{equation}
 P_l(x)=\frac{1}{2^l l!}\left(\frac{d}{dx}\right)^{l}(x^2-1)^l
\end{equation}

The solutions $|P_l^m(\cos\theta)|^2$ represent the probability of
finding the particle at a certain angle $\theta$, which means that
they give the probability distribution of the particle about the
$z$-axis. We will show below that we can view this distribution as
being shaped by a polar potential.


\section{The Polar Potential}
Now we can bring  the polar equation \eqref{thetaeq1}  to the
Schr\"{o}dinger form, just as we did with the radial
equation. Defining:
\begin{equation}\label{transtheta}
 \Theta(\theta)=\sin^{-\frac{1}{2}}(\theta)y(\theta)
\end{equation}

and multiplying by 1/2, equation \eqref{thetaeq1} becomes

\begin{equation}\label{thetasch1}
 -\frac{1}{2}\frac{d^2y(\theta)}{d\theta^2}+\left\{\frac{m^2-\frac{1}{4}}{2
\sin^2(\theta)}
\right\}y(\theta)=E y(\theta)
\end{equation}

where $E=(1/2)(l(l+1)+1/4)$.

\begin{figure}
\centering
    \includegraphics[width= \textwidth]{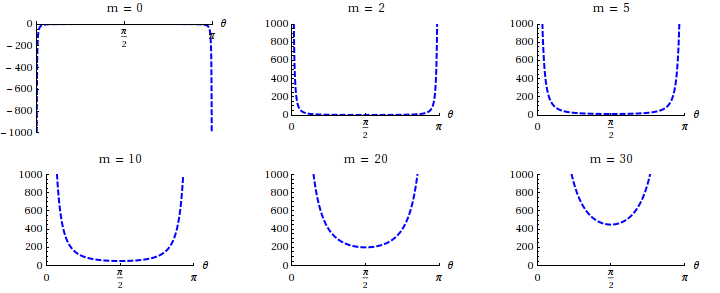}
\caption{Graphs of the potential for several values of integer m}
\label{potential}
\end{figure}

This equation can be thought of as a one dimensional
Shr\"{o}dinger equation (with $\hbar=m=1$ ) for a particle
confined between $0$ and $\pi$, and  satisfying the boundary
conditions $y(0)=y(\pi)=0$. The term in brackets is a one
dimensional potential, more precisely it is  a family of potentials
depending on the choice of the value of $m$. These potentials are
special cases of the so called P\"{o}schl-Teller
potential \cite{poschl,flugge} . To gain a closer look at the
potential, we plot it for various $m$-values in
Fig.\ref{potential}. As the figure intuitively suggests, the
potential, which is infinite at the boundaries - thus banning the
particle from going beyond the boundary- attempts to confine the
particle about the $z$-axis, the higher $m$ is the stronger is the
confining strength. For the lower $m$ values the potentials are
very close to an infinite well with width $\pi$, as we increase
$m$ the potential squeezes the states resulting in more
confinement.\\

\begin{figure}
\centering
    \includegraphics[width=.5 \textwidth]{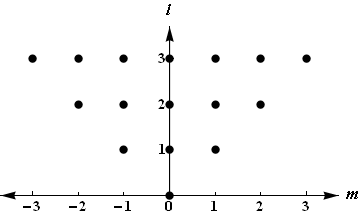}
\caption{$l$ and $m$ lattice}
\label{grid}
\end{figure}

 Let's take a look  at the solutions now. The solutions for
this family of potentials can be worked out in several ways, a
nice algebraic way is to use Infeld and Hull factorization method
\cite{infeldHull}. This family of potentials enjoys shape invariance symmetry, which can be exploited to workout the eigenvalues
algebraically. Supersymmetric quantum mechanics techniques can
also be used to work out the eigenstates \cite{dutt}. We will not
use these techniques here, rather we will work out the solutions
from the solutions of the original angular equation
\eqref{thetaeq1}. If we restrict the values of $m$ to nonnegative
integers, the solutions, upon employing the transformation
\eqref{transtheta} can be written directly from  the solutions of
the original equation and they read
\begin{equation}
 y_n^m(\theta)=N_n^m \sin^{\frac{1}{2}}\theta\ P_{n+m}^m(\cos\theta)
\end{equation}

where $P_{n+m}^m$ are, again, the associate Legendre functions.
These solutions are for nonnegative $m$, those for negative $m$
are - as is well-known \cite{griffithsQM,boas}- directly
proportional to these solutions. So, without loss of generality,
we will restrict ourselves to nonnegative $m$ values. In the
original equation \eqref{thetaeq1}, the solutions are
conventionally labeled by a pair of numbers ($m$,$l$) and subject
to the condition $m\leq l$. The pairs ($m$,$l$) constitute a
lattice as shown in Fig.\ref{grid}. In the conventional
interpretation of the solutions of \eqref{thetaeq1} the number $l$
is associated with the angular momentum, so the usual way to look
at the grid of solutions is to fix the angular momentum by fixing
$l$ and then changing $m$, so all the solutions with different
$m$'s and same $l$ are degenerate and correspond to the same
angular momentum. This amounts to moving right (or left) on the
grid. But what we are doing here is different, we are fixing $m$
and looking at the solutions of a one-dimensional potential at a
fixed $m$ value. When we fix the value of $m$ to a specific
integer we get all the solutions for that one dimensional
potential. As one can see, the first eigenstate for $m=1$ is at
$l=1$, but it still corresponds to the ground state of $m=1$
potential, which will be labeled as $n=0$, where $n$ is the
quantum number which labels the states of a fixed $m$ potential.
The same thing happens with all other $m$-values, thus we are
moving up the grid shown in Fig.\ref{grid}.  It is always the case
that the ground state of a fixed $m$ potential coincides with a
$l=m$, therefore, the relation between $l$ and $n$ is $l=m+n$.
Consequently, the eigenergies can now be written down directly
from those of equation \eqref{thetasch1}; they read:

\begin{equation}
\label{energies}
 E_n^m=\frac{1}{2}(n+m+\frac{1}{2})^2
\end{equation}

 Here, higher values of $n$, i.e. excited states correspond to
 higher values of the kinetic energy of the confined particle. That this corresponds
to higher $l$-values -in the conventional view- is natural, as
higher kinetic energies mean higher angular momenta.

\begin{figure}[h]
\centering
    \includegraphics[width= \textwidth]{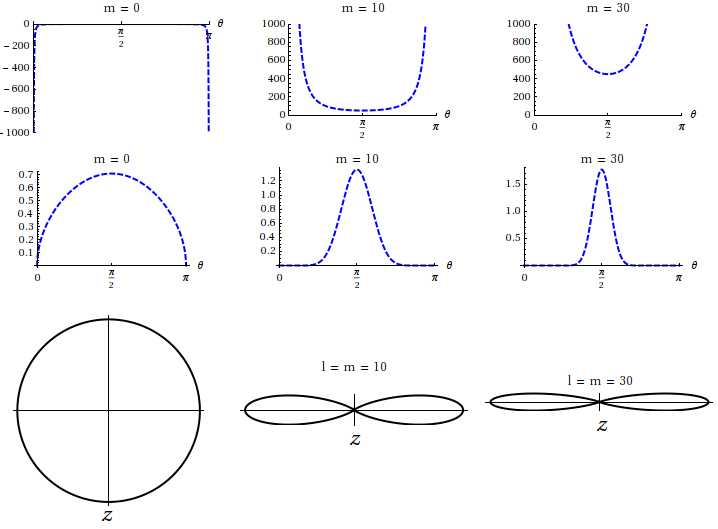}
\caption{The first row shows the potential for three values of
$m$, the second row shows the ground state solutions for those
potentials, the last line shows the polar plots for the Legendre
functions with the argument $\cos\theta$ for the corresponding
values of $m$ and $l$} \label{last}
\end{figure}

\begin{figure}[h]
\centering
    \includegraphics[width= \textwidth]{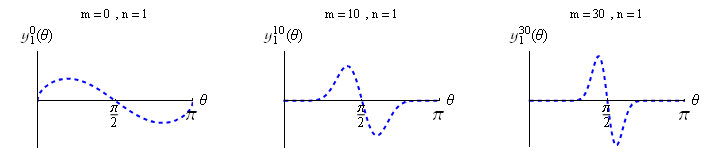}
\caption{The first excited states for $m=0,10,30$.The excited
states get more confined about $\theta=\pi/2$ as $m$ increases}
\label{firstExcited}
\end{figure}

Now, we show how to relate the behavior of the potential to that
of the eigenenergies, eigenstates and the original associate
Legendre functions. Let's look again at the graphs of the
potential for several $m$ in Fig.(\ref{potential}). Here we note
that  as $m$ increases the bottom of the potential is shifted up,
and this is responsible for increasing the ground state energy
with increasing $m$. The second observation here is that as one
goes for higher values of $m$, the potential tends to confine the
particle closer to the mid point of the potential
$\left(\theta=\frac{\pi}{2}\right)$, which means  confining the
particle closer to the $xy$-plane. This is confirmed by looking at
the behavior of the ground states for several $m$. In Fig.
(\ref{last}) we plot the ground states for various values of $m$
just below the corresponding potential.  It is clear that as $m$
increases the ground state gets squeezed closer to middle of the
box,
 giving rise to more confinement of the particle. The last line in
  Fig.(\ref{last}) shows the corresponding associated
Legendre functions which are related to the eigenfunctions through
the transformation \eqref{transtheta}. It is quite obvious that
the particle is more likely to be found near the $xy-$plane as $m$
increases. Since higher $m$ means higher angular momentum, this
squeezing of the states towards the $xy-$plane can be explained
semi-classically in terms of the increase of the moment of inertia
of the particle about the $z-$axis. Finally, while the examples we
have considered above are for the ground states only, the same
arguments apply equally well to the excited states. Fig.
(\ref{firstExcited}) shows the first excited states; $n=1$, for
three $m$-values. It is clear that as one goes up in $m$, the
particle gets more squeezed towards the $xy$-plane.

In conclusion, we see that we can view the angular distribution in
the wave functions of a particle in a central potential as being a
result of confinement by an $m-$dependent "effective" polar
potential.

\bibliographystyle{unsrt}
\bibliography{JJ2}
\end{document}